\documentclass[aps,prb,twocolumn,superscriptaddress]{revtex4}

\usepackage[dvips]{graphicx}
\usepackage{times,bm} 
\usepackage{amsmath}

\newcommand{\bQ}{\mbox{\boldmath $Q$}}

\newcommand{\bq}{\mbox{\boldmath $q$}}
\newcommand{\br}{\mbox{\boldmath $r$}}

\newcommand{\bS}{\mbox{\boldmath $S$}}

\begin{document}

\title{{\em Ab initio} Derivation of Low-Energy Model for Alkali-Cluster-Loaded Sodalites}

\author{Kazuma Nakamura} 
\thanks{Electronic mail: kazuma@solis.t.u-tokyo.ac.jp}
\affiliation{Department of Applied Physics, University of Tokyo, 7-3-1 Hongo Bunkyo-ku, Tokyo 113-8656, Japan} 
\affiliation{JST, CREST, 7-3-1 Hongo, Bunkyo-ku, Tokyo 113-8656, Japan}

\author{Takashi Koretsune} 
\affiliation{Department of Physics, Tokyo Institute of Technology, 2-12-1 Ookayama Meguro-ku, Tokyo 152-8551, Japan} 

\author{Ryotaro Arita} 
\affiliation{Department of Applied Physics, University of Tokyo, 7-3-1 Hongo Bunkyo-ku, Tokyo 113-8656, Japan}
\affiliation{JST, CREST, 7-3-1 Hongo, Bunkyo-ku, Tokyo 113-8656, Japan}
 
\date{\today}

\begin{abstract}
An effective low-energy model describing magnetic properties of alkali-cluster-loaded sodalites is derived by {\em ab initio} downfolding. 
We start with constructing an extended Hubbard model for maximally localized Wannier functions. 
{\em Ab initio} screened Coulomb and exchange interactions are calculated by constrained random phase approximation. 
We find that the system resides in the strong coupling regime and thus the Heisenberg model is derived as a low-energy model of the extended Hubbard model. 
We obtain antiferromagnetic couplings $\sim$ $O$(10 K), being consistent with the experimental temperature dependence of the spin susceptibility.
Importance of considering the screening effect in the derivation of the extended Hubbard model is discussed.

\end{abstract}


\maketitle
 
\section{Introduction}
Zeolites constitute a huge family of nonporous crystalline aluminosilicates which have a wide variety of intriguing properties.\cite{Zeolite_Book}
Because of their capability of hosting various ions, atoms, molecules and clusters in their subnanometric pores with rich possibilities of crystal structure, they have versatile functionalities such as high catalytic activities, sorption characteristics, ion-exchange abilities.
Numerous zeolites with various compositions and framework topologies have been synthesized and exploited in many applications. 
Besides such diverse fascinating aspects, it is of great interest to focus on electron correlations in this system. 
A variety of zeolites have been viewed as three-dimensional correlated {\it s}-electron systems providing a non-trivial play ground for a systematic control of many-body correlation effects.\cite{Arita} 
For example, although all the ingredients of zeolites are nonmagnetic elements, some of zeolites exhibit an intriguing magnetism for certain conditions; zeolites LTA and LSX with potassium clusters have ferromagnetic ground states depending on the number of potassium atoms per cage,\cite{Nozue_LTA,Nozue_LSX,Nakano} while sodalites loading various alkali-metal clusters exhibit robust antiferromagnetism.\cite{Srdanov,NMR,muSR,Nozue_SOD}

When we study such characteristic many-body effects in zeolitic materials, it is definitely impractical to calculate everything from first principles.
The unit cell is extremely huge and contains many atoms, so that formidable numerical cost would be required. On the other hand, recent conventional {\it ab initio} studies have clarified that some zeolites have quite simple low-energy electronic structures;\cite{Arita,Blake,Madsen,Windiks,Sankey} 
for example, in the sodalite system, the aluminosilicate cage forms a wide gap more than 5 eV around the Fermi level and, in this energy gap, electronic states due to guest alkali clusters make narrow bands with the width $\sim$1 eV.
Its band dispersion is well represented by simple tight-binding models, which indicates that the so-called ``superatom" picture\cite{Arita} or the ``particle-in-a-box" model\cite{Sankey} correctly captures essential aspects of the low-energy physics of the sodalite systems.
With this situation, rather than the full {\it ab initio} approach, the three-stage approach is expected to work more successfully. 
Here, in the first stage, we perform standard {\it ab initio} density-functional calculations and, in the next step, we perform downfolding procedures, that is, construction of an effective low-energy model. 
Finally, we solve the resulting model by high accurate and reliable solvers. 
The so-called ``LDA+DMFT" method\cite{LDA+DMFT} combining local density approximation (LDA) and dynamical mean field theory (DMFT) is a typical example of this approach. 
Recently, the three-stage approach has been extensively applied to various correlated electron systems. 
Especially, it has been demonstrated that the scheme really works with high accuracy for various transition-metal oxides.\cite{LDA+DMFT,PIRG} 

Recently, as a reliable tool for evaluating the values of interaction parameters in the downfolding step, a constrained random phase approximation (cRPA) method is formulated.\cite{Aryasetiawan,Solovyev} 
Compared to the standard method based on a constrained LDA technique,\cite{cLDA} the cRPA has several advantages; one can precisely exclude screening processes between the basis states of the effective model, which should be considered in the last stage solving the effective model.
In addition, we can obtain onsite and offsite interactions at one time.
While the cRPA method has been employed in many studies,\cite{PIRG,Ferdi,Miyake,Nakamura} applications to zeolitic materials have yet to be done. 
Indeed, it is a highly non-trivial issue to determine the values of interaction parameters of zeolites; the bases of the low-energy model of these materials are no longer localized at some specific atoms and are extended spatially beyond several guest atoms in the cage.
So, we have to evaluate the value of interaction parameters not for atomic orbitals but for molecular orbitals. 
In order to construct automatically such basis functions with non-trivial spatial spread, it is convenient to exploit maximally localized Wannier orbitals (MLWOs). \cite{Ref_MLWF}
Recently, MLWO is combined with cRPA calculations to estimate the onsite Hubbard $U$ as well as offsite interaction parameters in the low-energy models of various systems.\cite{Miyake,Nakamura} 

It is of great interest to apply this state-of-the-art downfolding technique based on the combination of cRPA with MLWO to zeolitic materials and examine how it works. 
As a benchmark for this purpose, we consider sodalites which are classified as the simplest zeolite. 
The framework of this material is described as a body-centered-cubic (bcc) array of $\beta$ cages [(SiO$_2$)$_3$(AlO$_2^-$)$_3$, the smallest unit of the aluminosilicate cage] and each cage accommodates ionic alkali clusters A$_4^{3+}$ to neutralize the negative charge of the framework. 
Experimentally, it has been well established that the system can be viewed as a crystal of $F$ centers sitting on the center of the tetragonal cluster A$_4^{3+}$. 
Especially, a magnetic property of a sodium electrosodalite (or black sodalite) is quite well understood in terms of the $S$=1/2 Heisenberg model on the bcc lattice.\cite{Srdanov,NMR,muSR,Nozue_SOD}
 Since the temperature dependence of the magnetic susceptibility of the Heisenberg model can be calculated by the high-temperature-expansion scheme with high accuracy, we can obtain the precise values of exchange coupling by parameter fitting to the experimental data.
Thus, the sodalites are the best systems to examine the reliability of derived parameters with the {\em ab initio} downfolding, in that we can compare unambiguously the theoretical exchange values and the experimental ones.

The purpose of the present study is to examine how accurately we can construct a low-energy model of the sodalite system by the {\it ab initio} downfolding technique.
While one can exploit direct {\em ab initio} calculations based on local spin density approximation to evaluate the exchange coupling,\cite{Oguchi} the present study focuses on an alternative approach which is feasible to not only localized spin systems but also more general cases. 
First, we construct a single-band extended Hubbard model based on the {\it ab initio} downfolding scheme and then derive an effective Heisenberg model by the second-order perturbation. 
A similar strategy was taken in the previous study\cite{Sankey} but, there, the ``kinetic-exchange" term\cite{Anderson} only was evaluated and the ``direct-exchange" term was completely neglected. 
As shown below, in the sodalite, the direct exchange has the same energy scale as the kinetic exchange and thus the two exchange couplings compete with each other. 
In addition, in the past parameter estimations, the screening effect was completely neglected.
We will show the importance of taking the screening effect into account in the parameter derivation; if we neglect the screening effect, the kinetic-exchange value is smaller than the direct-exchange value and the net exchange becomes ferromagnetic. 
When the screening effect is switched on, the kinetic exchange reverses the direct exchange, thus resulting in antiferromagnetic interactions being consistent with the experiments.

This paper is organized as follows.
In Sec.~II, we describe our basic strategy for deriving the effective Heisenberg model from first principles. 
Section~III is devoted to the accurate estimation of the experimental exchanges using the high-temperature expansion to the Heisenberg model. 
Following by recent measurement of the magnetic susceptibility for the sodalites,\cite{Nozue_SOD} we give the exchange parameters of the sodium electrosodalite and the potassium electrosodalite and discuss the differences between the two. 
In Sec.~IV, {\em ab initio} computational results are presented and compared with the experimental results. 
The concluding remarks are given in Sec.~V.     

\section{{\em Ab initio} construction of effective Hamiltonians}
We consider {\em ab initio} derivations of the effective Heisenberg model describing ``low-energy" electronic structures. Conventionally, the derivation is based on the second-order perturbation to the single-band extended Hubbard Hamiltonian consisting of the transfer part $\mathcal{H}_{t}$, the Coulomb-interaction part $\mathcal{H}_{V}$, and the exchange-interaction part $\mathcal{H}_{J}$ as  
\begin{eqnarray}
\mathcal{H}=\mathcal{H}_t + \mathcal{H}_{V} + \mathcal{H}_{J} \label{H_Hub}
\end{eqnarray} 
with 
\begin{eqnarray}
\mathcal{H}_t = \sum_{\sigma} \sum_{ij} t_{ij} 
a_{i \sigma}^{\dagger} a_{j \sigma},  \label{H_t}
\end{eqnarray} 
\begin{eqnarray}  
\mathcal{H}_{V} = \frac{1}{2} \sum_{\sigma \rho} \sum_{ij} 
 V_{ij} a_{i \sigma}^{\dagger} a_{j \rho}^{\dagger}
        a_{j \rho} a_{i \sigma}, \label{H_U}
\end{eqnarray} 
\begin{eqnarray}  
\mathcal{H}_{J} = \frac{1}{2} \sum_{\sigma \rho} \sum_{ij} 
 J_{ij} a_{i \sigma}^{\dagger} a_{j \rho}^{\dagger}
        a_{i \rho} a_{j \sigma},  \label{H_J}
\label{eq:H}                
\end{eqnarray}
where $a_{i \sigma}^{\dagger}$ ($a_{i \sigma}$) is a creation (annihilation) operator of an electron with spin $\sigma$ in the Wannier orbital localized in the $i$th sodalite cage. 
The $t_{ij}$\ parameters in Eq.~(\ref{H_t}) contain an onsite energy ($i$ = $j$) and hopping integrals ($i \ne j$), written by 
\begin{eqnarray}
t_{ij} = \langle \phi_{i} | \mathcal{H}_0 |  \phi_{j} \rangle \label{t_ij} 
\end{eqnarray}
with $| \phi_{i} \rangle =a_{i}^{\dagger}|0\rangle$
and $\mathcal{H}_0$ being the one-body part of $\mathcal{H}$.
The $V_{ij}$ and $J_{ij}$ values in Eqs.~(\ref{H_U}) and (\ref{H_J}) are screened Coulomb and exchange integrals in the Wannier orbital, respectively, expressed as 
\begin{widetext}  
\begin{eqnarray}
V_{ij}
= \langle \phi_{i} \phi_{j} | W | \phi_{i} \phi_{j} \rangle 
=\int \int d\br d\br' \phi_{i}^{*}(\br) \phi_{i}(\br) 
W(\br,\br') \phi_{j}^{*}(\br') \phi_{j}(\br') \label{V_ij}
\end{eqnarray} 
and 
\begin{eqnarray}
J_{ij} 
= \langle \phi_{i} \phi_{j} | W | \phi_{j} \phi_{i} \rangle
= \int \int d\br d\br' \phi_{i}^{*}(\br) \phi_{j}(\br) 
W(\br,\br') \phi_{j}^{*}(\br') \phi_{i}(\br'), \label{J_ij}
\end{eqnarray}
\end{widetext}  
where $W(\br,\br')$ is a screened Coulomb interaction. 
$V_{ij}$ at $i$ = $j$ corresponds to onsite Hubbard parameter $U$. 

Now, we consider a situation with the half-filling and atomic-limit condition, where the parameters satisfy the following inequality
\begin{eqnarray}
U - V_{ij} \gg |t_{ij}| > 0.
\end{eqnarray}
In this situation, with the second-order perturbation, the effective Hamiltonian which describes the fine ``low-energy" spectrum associated with the spin structure is given as the following Heisenberg model\cite{Yoshida} 
\begin{eqnarray}
\mathcal{H}_{{\rm eff}} = 2 \sum_{i > j} \mathcal{J}_{ij} \bS_{i} \cdot \bS_{j}, \label{H_eff}
\end{eqnarray}
where the local spin operator $\bS_{i}$ is conventionally represented in term of the creation and annihilation operators as $S_{i}^{x}=\frac{1}{2}(a_{i \uparrow}^{\dagger} a_{i \downarrow} + a_{i \downarrow}^{\dagger} a_{i \uparrow})$, $S_{i}^{y}=\frac{1}{2i}(a_{i \uparrow}^{\dagger} a_{i \downarrow} - a_{i \downarrow}^{\dagger} a_{i \uparrow})$, and $S_{i}^{z}=\frac{1}{2}(a_{i \uparrow}^{\dagger} a_{i \uparrow} - a_{i \downarrow}^{\dagger} a_{i \downarrow})$. 
The effective exchange coupling in Eq.~(\ref{H_eff}) is written as 
\begin{eqnarray}
\mathcal{J}_{ij}= K_{ij} - J_{ij} \label{J_eff}
\end{eqnarray}
with 
\begin{eqnarray}
K_{ij} = \frac{ 2 |t_{ij}|^{2} } { U - V_{ij} }. \label{K_ij}
\end{eqnarray}
The $K_{ij}$ is a ``kinetic-exchange" term \cite{Anderson} which stabilizes the antiferromagnetic coupling between the local spins, while the second term in Eq.~(\ref{J_eff}) is a ``direct-exchange" term favoring the ferromagnetic coupling.  
The competition between the two-type exchange terms determines the net magnetic feature of the system (i.e., whether the system prefers the antiferromagnetic or ferromagnetic state). 

The calculation of the effective exchanges $\mathcal{J}_{ij}$ is basically straightforward after parameterizations of $t_{ij}$, $V_{ij}$, and $J_{ij}$ but a careful treatment is needed for the calculation of the screened Coulomb interaction of $W(\br,\br')$.
The screened interaction considered in the extended Hubbard model should not include screening formed in a target band of the model. 
This screening should be considered at the step of solving the effective model and, at the downfolding stage, we must exclude the target-band screening effects to avoid the double counting of this screening.  
In the random phase approximation (RPA), this constraint is easily imposed,\cite{Aryasetiawan,Solovyev} because the RPA polarization function is given as the sum of the band pairs associated with individual transitions; we first calculate the polarization function with excluding the transitions in the target band and then evaluate the screened interaction $W(\br,\br')$ with using this polarization function.  
Finally, we compute the $V_{ij}$ and $J_{ij}$ parameters as the Wannier matrix elements of the $W$ interaction. 

There are two other choices on the treatment of the Coulomb interaction. 
The first is the use of ``bare" Coulomb interaction $v(\br,\br')$=$\frac{1}{|\br-\br'|}$ instead of $W(\br,\br')$. 
The resulting $V_{ij}$ and $J_{ij}$ parameters have no screening effect and will give larger values than the constrained-RPA values discussed above.
The kinetic-exchange parameter $K_{ij}$ becomes small because of the increase of $U-V_{ij}$ in Eq.~(\ref{K_ij}).
We note that this choice has been widely used in the literature so far \cite{Blake,Madsen,Windiks,Sankey} but there is no justification. 
Another choice is the use of the ``fully" screened Coulomb interaction, where we calculate the RPA polarization function with no constraint on the transitions. The result includes the target-band screening effect and therefore the calculated $V_{ij}$ and $J_{ij}$ values will be largely reduced compared to the constrained-RPA values. 
We compute the interaction parameters $V_{ij}$ and $J_{ij}$ with the ``bare", ``constrained RPA", and ``full RPA" interactions and discuss the importance of the screening effect on the derivation for exchange values of the Heisenberg model.
\section{Estimation of the exchange couplings from the experiment}
Before presenting {\em ab initio} computational results, we consider experimental values of exchange couplings, which are estimated from the data for the temperature dependence of the magnetic susceptibility. 
In the sodium electrosodalite, the measured Weiss temperature $\Theta_{\rm W}$ is $-$170 K, while the N\'eel temperature $T_{\rm N}$ is 50 K. 
The negative Weiss temperature and the existence of the antiferromagnetic transition indicate the antiferromagnetic interaction between neighboring spins, while inequality $|\Theta_{\rm W}|$$\gg$$T_{\rm N}$ implies that there is strong frustration in the system or equivalently the presence of next-nearest-neighbor exchange couplings.
In fact, magnetic properties of the sodium electrosodalite have been discussed with the Heisenberg model up to the next nearest neighbors. 
Recently, the magnetic measurement has been performed for the potassium electrosodalite and the $\Theta_{\rm W}$ and $T_{\rm N}$ temperatures are observed as $-$330 K and 80 K, respectively.\cite{Nozue_SOD} 
Here, we determine the exchange parameters in the Heisenberg model so that the calculated model Weiss and N\'eel temperatures reproduce the experimental ones.
Accuracy of $\Theta_{\rm W}$ and $T_{\rm N}$ obtained from solving the model critically affects the quality of the exchange couplings.
In this work, we calculate the high-temperature series of the spin susceptibility up to tenth order in inverse temperature \cite{HTEseries} using the finite cluster method.\cite{domb_green} 

The explicit form of the Heisenberg model on the bcc lattice up to the next nearest neighbors is given as
\begin{align}
{\mathcal H}_{{\rm eff}} = 2 {\mathcal J}_1 \sum_{\langle ij \rangle} \bS_i \cdot \bS_j + 2 {\mathcal J}_2 \sum_{\langle ij \rangle'} \bS_i \cdot \bS_j,
\end{align}
where the first summation ${\langle ij \rangle}$ is taken over the bonds between nearest neighbors and the second summation ${\langle ij \rangle}^\prime$ over the bonds between next-nearest neighbors.
${\mathcal J}_1$ and ${\mathcal J}_2$ represent the exchange couplings for the nearest neighbors and next-nearest neighbors, respectively.
Note that the suffices ``1" and ``2" attached to ${\mathcal J}$ hereafter specify the bond between the nearest neighbors and the bond between the next-nearest  neighbors, respectively. 

The spin susceptibility for a general wavevector, $\chi(\bq)$, can be expressed as \cite{Yoshida} 
\begin{align}
\chi(\bq) = \frac{1}{N} \int^{\beta}_0 d\tau \sum_{ij}
\langle e^{  {\mathcal H}_{{\rm eff}} \tau} S_i^z 
        e^{- {\mathcal H}_{{\rm eff}} \tau} S_j^z \rangle 
        e^{i \boldsymbol{q} \cdot (\boldsymbol{r}_i - \boldsymbol{r}_j)}.
\end{align}
Here, $\beta$ is the inverse temperature and $\langle \cdots \rangle$ represents the thermodynamic average; i.e., 
 $\langle \cdots \rangle = {\rm Tr}(\cdots e^{-\beta {\mathcal H}_{{\rm eff}}})  / {\rm Tr} ( e^{-\beta {\mathcal H}_{{\rm eff}}})$.
Uniform and staggered spin susceptibilities are given as $\chi$ = $\chi({\bf 0})$ and $\chi(\bQ)$ with $\bQ = (\pi,\pi,\pi)$, respectively. 
The $\chi$ and $\chi(\bQ)$ up to the first order in $\beta$ are given as
\begin{align}
4 \chi T &= 1 - \beta (4 {\mathcal J}_1 + 3 {\mathcal J}_2 ) + O(\beta^2),\label{eq:chi}\\
4 \chi(\bQ) T &= 1 + \beta ( 4 {\mathcal J}_1 - 3 {\mathcal J}_2 ) + O(\beta^2). \label{eq:chiq}
\end{align}
The first-order coefficients $-$(4${\mathcal J}_{1}$+3${\mathcal J}_{2}$) and 4${\mathcal J}_{1}$$-$3${\mathcal J}_{2}$ above correspond to the high-temperature-limit Weiss temperature\cite{dalton_wood} and the mean-field N\'eel temperature,\cite{rushbrooke_wood} respectively.
It should be noted here that the temperature range for experimental $\Theta_{\rm W}$ is far from the high-temperature limit, so that the Weiss temperature given above is not a good estimate for the experimental value. 
Furthermore, the mean-field N\'eel temperature is seriously overestimated because the quantum fluctuation is neglected.

\begin{figure}
\includegraphics[width=0.4\textwidth]{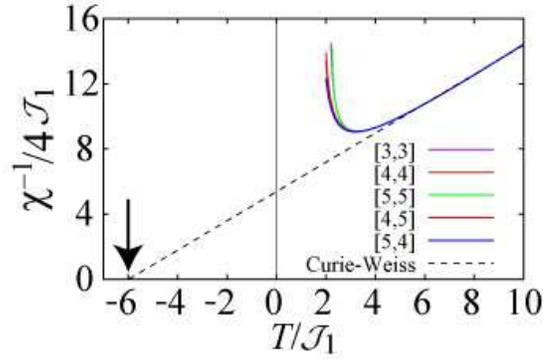} 
\caption{(Color online) Inverse of the uniform spin susceptibility at ${\mathcal J}_2 = 0$. $[3,3]$, $[4,4]$, $[5,5]$, $[4,5]$, and $[5,4]$ Pad\'e approximations are shown. We also plot the Curie-Weiss fitting as the dotted line. An arrow indicates the extrapolated Weiss temperature estimated as $-$6.0${\mathcal J}_1$. Notice that this value is largely deviated from the high-temperature-limit value $-$4${\mathcal J}_1$.}
\label{fig:chi}
\end{figure}

\begin{figure}
\includegraphics[width=0.4\textwidth]{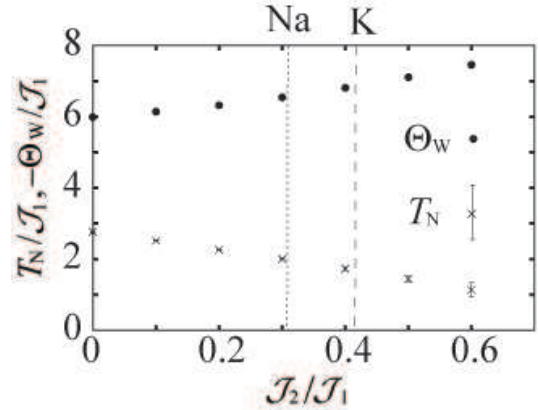} 
\caption{Weiss temperature, $\Theta_{\rm W}$, and N\'eel temperature, $T_{\rm N}$, as a function of ${\mathcal J}_2/{\mathcal J}_1$. Vertical errorbars for $T_{\rm N}$ represent the scattering of the various Pad\'e approximation. 
For sodium electrosodalite, the experimental $\Theta_{\rm W}$ ($-$170 K) and $T_{\rm N}$ (50 K) are well reproduced, when $\mathcal{J}_1$ and $\mathcal{J}_2/\mathcal{J}_1$ are set to 26 K and 0.31 (vertical dotted line), respectively.
 In the case of potassium electrosodalite, such $\mathcal{J}_1$ and $\mathcal{J}_2/\mathcal{J}_1$ were found to be 48 K and 0.42 (vertical dashed line), respectively.}
\label{fig:tntw}
\end{figure}

To go beyond the first-order analysis and obtain the precise temperature dependence of the magnetic susceptibility, we consider the higher-order expansion series of $\chi$ and $\chi(\bQ)$.
In order to extrapolate the series down to low temperatures, we use the Pad\'e approximation, in which a series of $\chi$ and $\chi(\bQ)$ is approximated as
$\frac{P_L(x)}{Q_M(x)}$,
where $P_L(x)$ and $Q_M(x)$ are the $L$-order and $M$-order polynomials, respectively. 
We call it the $[L,M]$ Pad\'e approximation.
Figure \ref{fig:chi} illustrates an example of the extrapolation of $\chi^{-1}$ at ${\mathcal J}_2 = 0$.
We find that various Pad\'e approximations show good convergence down to $T/{\mathcal J}_1$$\sim$3.
To estimate the Weiss temperature, $\Theta_{\rm W}$, we fit $\chi$ as $\chi = (T-\Theta_{\rm W})^{-1}$ as shown in the dotted line in Fig.\ \ref{fig:chi}.
The fitting temperature range is 5$<$$T/{\mathcal J}_1$$<$10. 
(As shown below, this range roughly corresponds to the experimental temperature range.) 
The $\Theta_{\rm W}$ estimated as $-$6.0${\mathcal J}_1$ indicated by the arrow in the figure is appreciably smaller than the first-order value $-$4${\mathcal J}_1$.
The treatment can straightforwardly be applied to the case of ${\mathcal J}_2$$\ne$0. 
In Fig.\ \ref{fig:tntw}, the calculated Weiss temperature (dots) is shown as a function of ${\mathcal J}_2/{\mathcal J}_1$.


The estimation for $T_{\rm N}$ proceeds as follows: $\chi(\bQ)$ is expected to have a pole at the finite $\beta$ and behaves as
\begin{align}
\chi(\bQ) \propto (\beta - \beta_{\rm N})^{-\gamma}, \label{chiq_beta}
\end{align}
where $\beta_{\rm N}$ is the inverse of the antiferromagnetic transition temperature ($T_{\rm N} \equiv 1/\beta_{\rm N}$) and the $\gamma$ is the critical exponent.
In the three-dimensional Heisenberg model, $\gamma$ is known to be $\sim$1.39 (Ref.~\onlinecite{guillou_zinnjustin}).
By taking the logarithmic derivative of Eq.~(\ref{chiq_beta}), we obtain\cite{domb_green} 
\begin{align}
\frac{d \log \chi(\bQ)}{d \beta} \propto \frac{-\gamma}{\beta - \beta_{\rm N}}.
\end{align}
Since the Pad\'e approximation can describe simple poles exactly, approximations to the logarithmic derivative should converge much faster.
In addition, we can evaluate $\gamma$ from a residue of the pole as well as the location of the pole giving the critical temperature. 
At ${\mathcal J}_2 = 0$, various Pad\'e approximations show good convergence and give $T_{\rm N} = 2.76$ and $\gamma = 1.39$.
For finite ${\mathcal J}_2$, however, the frustration lowers the N\'eel temperature and it becomes difficult to estimate $T_{\rm N}$ and $\gamma$ accurately.
Thus, to improve the convergence, we use the Pad\'e approximation of $\chi^{1/\gamma}$ with $\gamma$ kept at 1.39.
This assumption works well even for the finite ${\mathcal J}_2$ and we obtain $T_{\rm N}$ as a function of ${\mathcal J}_2$/${\mathcal J}_1$ (crosses of Fig.\ \ref{fig:tntw}).
The errorbar comes from the scattering of the various Pad\'e approximations.
By using these data and referring the experimental $\Theta_{\rm W}$ and $T_{\rm N}$ temperatures, we reasonably estimate the exchange couplings of the sodalite system. 
The resulting values are 
${\mathcal J}_1$ = 26 K and ${\mathcal J}_2$ = 8 K for the sodium electrosodalite and 
${\mathcal J}_1$ = 48 K and ${\mathcal J}_2$ = 20 K for the potassium electrosodalite.
 
\section{Results and Discussions}
Our {\em ab initio} calculations were performed with {\em Tokyo Ab initio Program Package}.\cite{Ref_TAPP} 
With this program, electronic-structure calculations with the generalized-gradient-approximation (GGA) exchange-correlation functional \cite{Ref_PBE96} were performed using a plane-wave basis set and the Troullier-Martins norm-conserving pseudopotentials \cite{Ref_PP1} in the Kleinman-Bylander representation. \cite{Ref_PP2,Ref_PP3}
The energy cutoff in the band calculation was set to 49 Ry and a 5$\times$5$\times$5 $k$-point sampling was employed. 
The experimental crystal-structure data were taken from Ref.~\onlinecite{Sankey} for sodium electrosodalite and Ref.~\onlinecite{Madsen} for potassium electrosodalite.  
The calculations for the screened interactions are followed by Ref.~\onlinecite{Nakamura}. 
The polarization function was expanded in plane waves with an energy cutoff of 5 Ry and the total number of bands considered in the polarization calculation was set to 200. 
The Brillouin-zone integral on wavevector was evaluated by the generalized tetrahedron method.\cite{Fujiwara} 
The additional terms in the long-wavelength polarization function due to nonlocal terms in the pseudopotentials were explicitly considered following Ref.~\onlinecite{Louie}. 
A problem due to the singularity in the Coulomb interaction, in the evaluation of the Wannier matrix elements $V_{ij}$ and $J_{ij}$, was treated in the manner described in Ref. \onlinecite{Louie}.

\begin{figure}[h]
\includegraphics[width=0.3\textwidth]{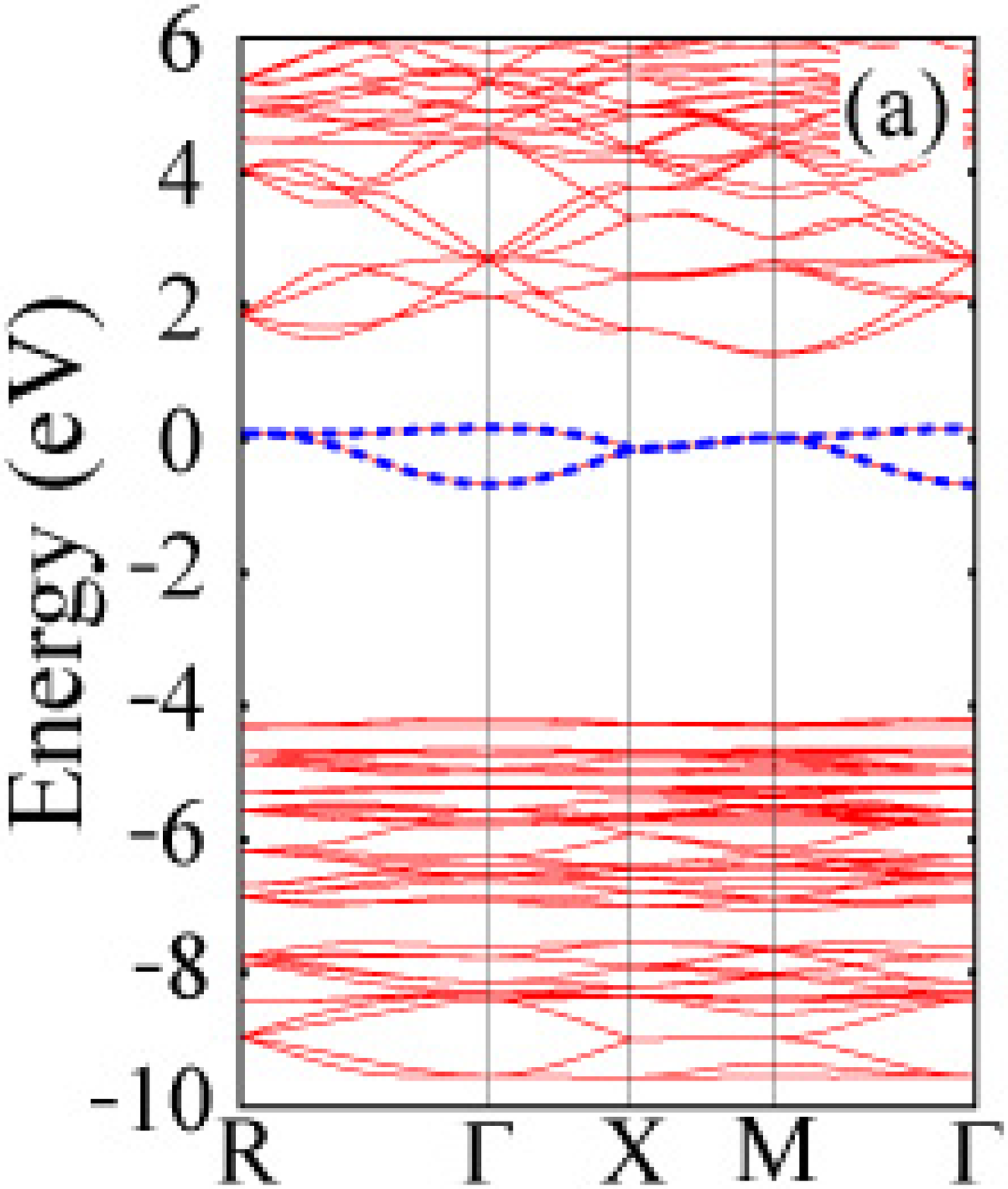} 
\includegraphics[width=0.3\textwidth]{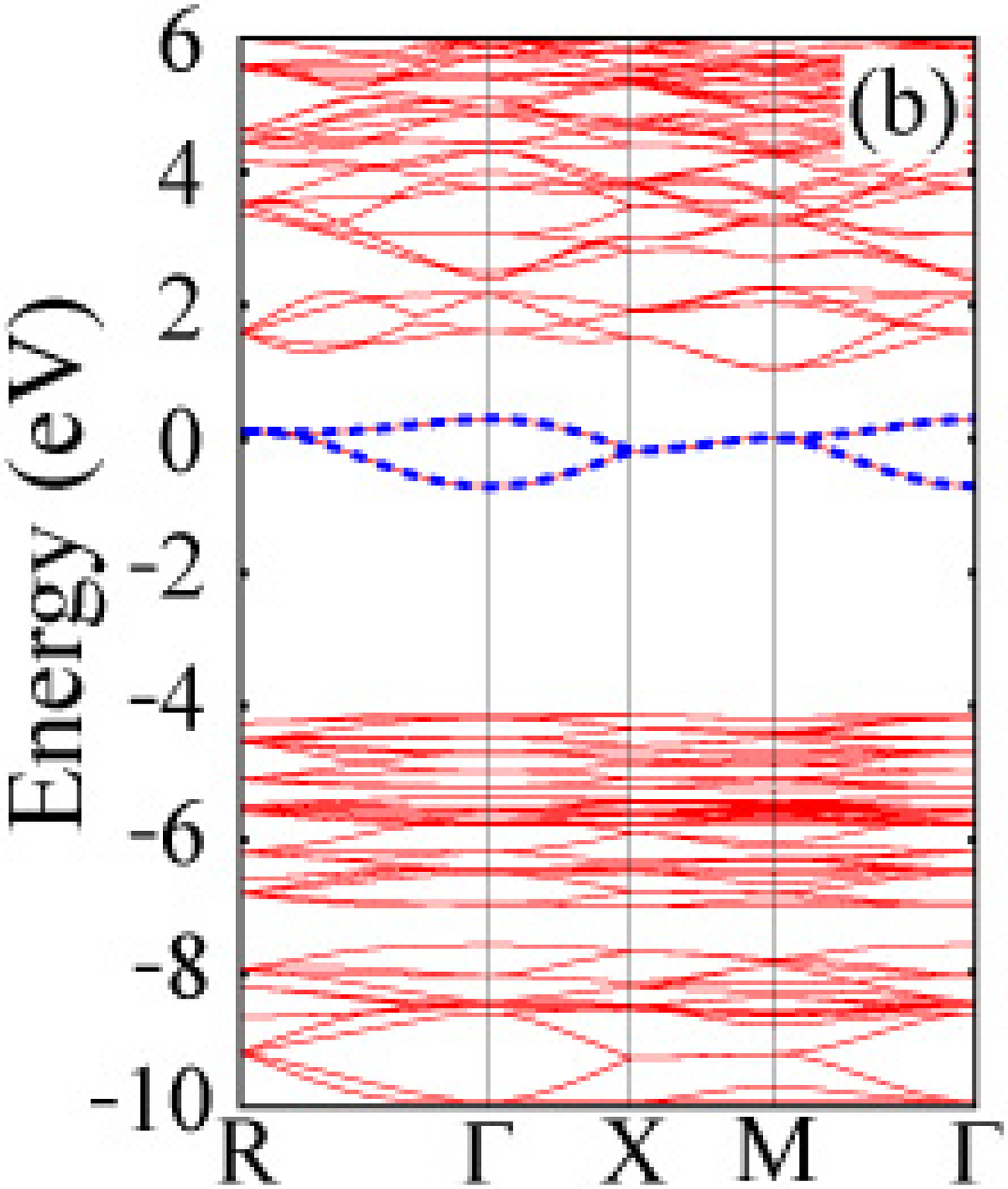}
\caption{(Color online) Calculated {\em ab initio} band structures (red solid lines) of (a) sodium electrosodalite and (b) potassium electrosodalite. The blue dotted dispersions are obtained by the $t_1$-$t_2$ model, where $t_1$ and $t_2$ are nearest and next-nearest transfers, respectively. For the values, see the text. The zero of energy is the Fermi level.}
\label{fig:band}
\end{figure}

We show in Fig.~\ref{fig:band} {\em ab initio} GGA band structures (red solid lines) of (a) sodium electrosodalite and (b) potassium electrosodalite.
We see an isolated band near the Fermi level (energy zero).
This band is due to confined electrons in the sodalite cage and we employ this band as the target band of the extended Hubbard model.
The entangled band structures below $-$4 eV and above $+$1 eV are associated with electronic states of the framework of the sodalite.  
The overall band structure of the sodium electrosodalite is similar to that of the potassium electrosodalite.
A notable difference is that 
the target bandwidth of the sodium electrosodalite is 0.86 eV, while that of the potassium electrosodalite is 1.01 eV, which makes differences in the values of transfer integrals of the two materials (see below).

\begin{figure}[h]
\includegraphics[width=0.30\textwidth]{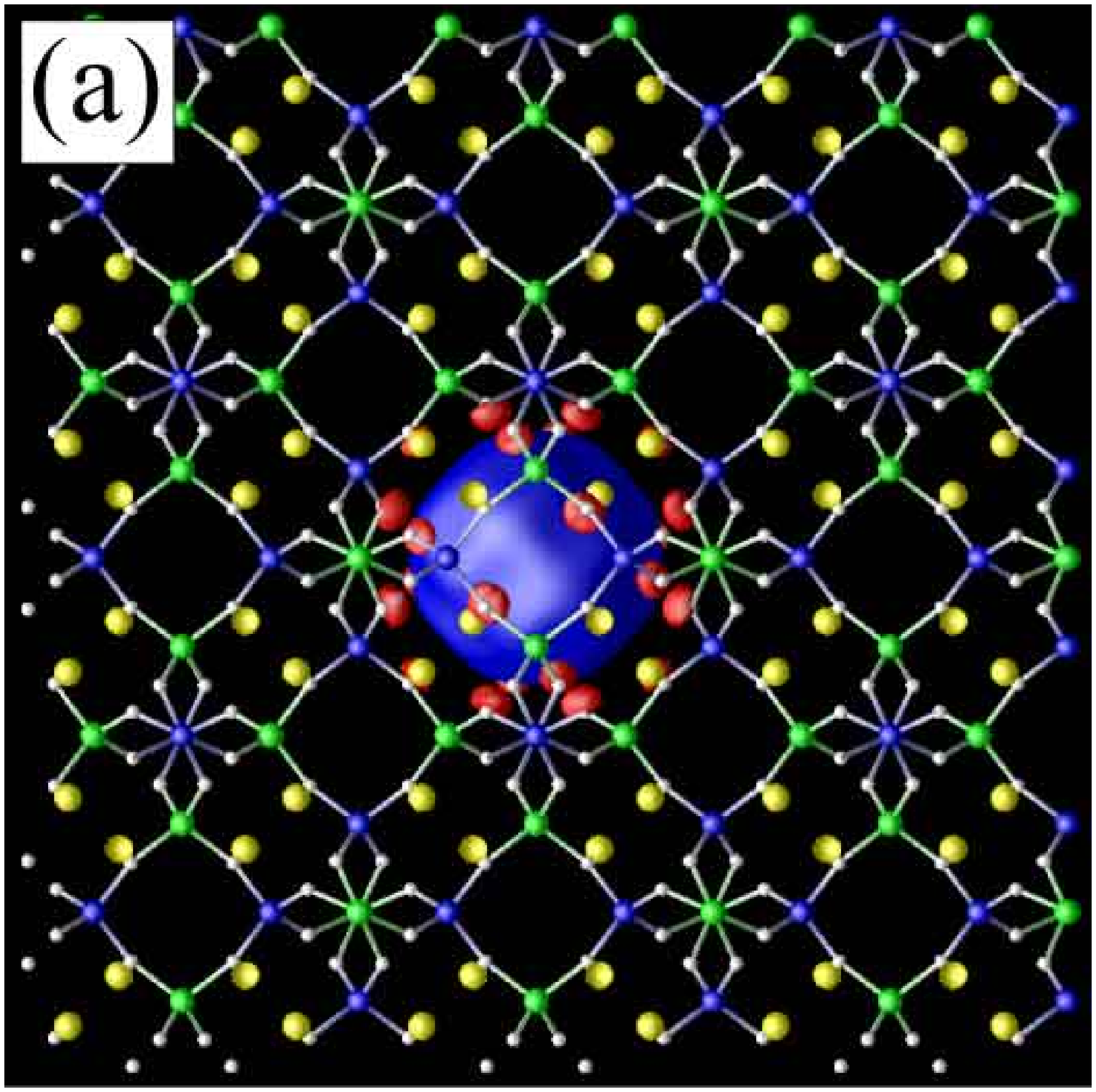} 
\includegraphics[width=0.30\textwidth]{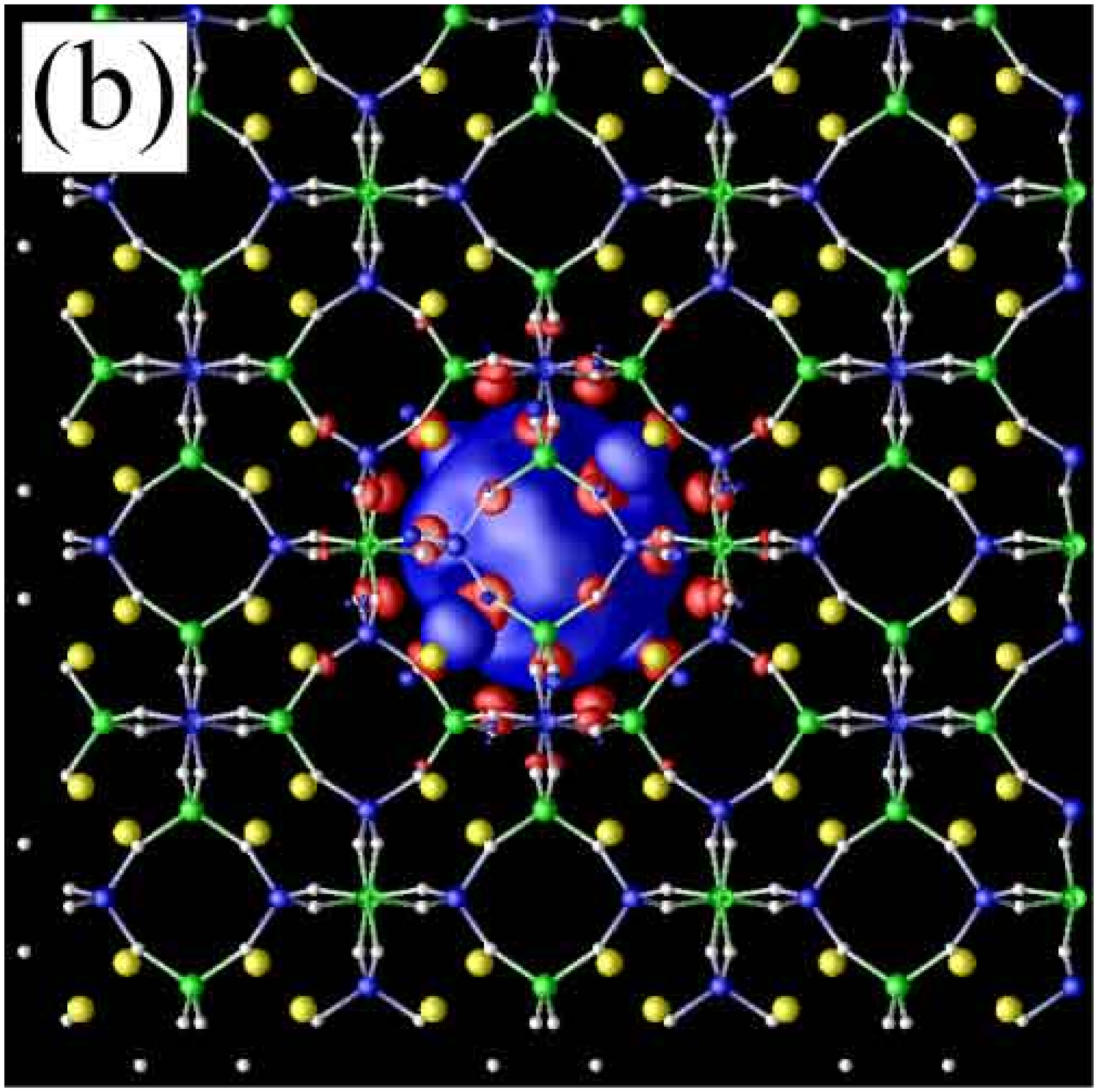} 
\caption{(Color online) Calculated maximally localized Wannier functions of (a) sodium electrosodalite and (b) potassium electrosodalite. The amplitudes of the contour surface are +1.5/$\sqrt{v}$ (blue) and $-$1.5/$\sqrt{v}$ (red), where $v$ is the volume of the primitive cell. Si, O, Al, and Na or K nuclei are illustrated by blue, silver, yellow, and green spheres, respectively. } 
\label{fig:Wannier}
\end{figure}

Figure~\ref{fig:Wannier} visualizes our calculated maximally localized Wannier orbitals for the target band of (a) sodium electrosodalite and (b) potassium electrosodalite.
We can see that the resulting Wannier orbital is confined in the cage and has an $s$ symmetry around the cage center. 
The calculated spatial spread of the Wannier orbitals are 2.66 \AA\ for the sodium electrosodalite and 2.91 \AA\ for the potassium electrosodalite and these values are smaller than the diameter of the cage (7.6 \AA\ for the sodium electrosodalite and 8.0 \AA\ for the potassium electrosodalite). 

We next calculate transfer integrals in Eq.~(\ref{t_ij}) as matrix elements of the Kohn-Sham Hamiltonian in the Wannier orbital.
The nearest-neighbor transfer $t_1$ and the next-nearest-neighbor transfer $t_2$ are found to be $-$57.3 meV and $-$32.1 meV for the sodium electrosodalite. 
The results for the potassium electrosodalite are $-$68.0 meV and $-$31.1 meV.
It was found to be negligibly small for other transfers beyond the third neighbors; their magnitudes are less than a few meV.   
The band dispersion calculated with $t_1$ and $t_2$ is shown as blue dots in Fig.~\ref{fig:band}.
We can see that the original band structure is quite well reproduced with the two-parameter model.
We note that the Kohn-Sham Hamiltonian $\mathcal{H}_{{\rm KS}}$ is different from the exact one-body Hamiltonian $\mathcal{H}_{0}$ in Eq.~(\ref{t_ij}).
The difference between the two requires involved discussions about the ``downfolding self energy",\cite{Aryasetiawan,Solovyev} so, in the present study, for the simplicity, we employed the $\mathcal{H}_{{\rm KS}}$ instead of the $\mathcal{H}_{0}$.

\begin{figure}[t]
\includegraphics[width=0.35\textwidth]{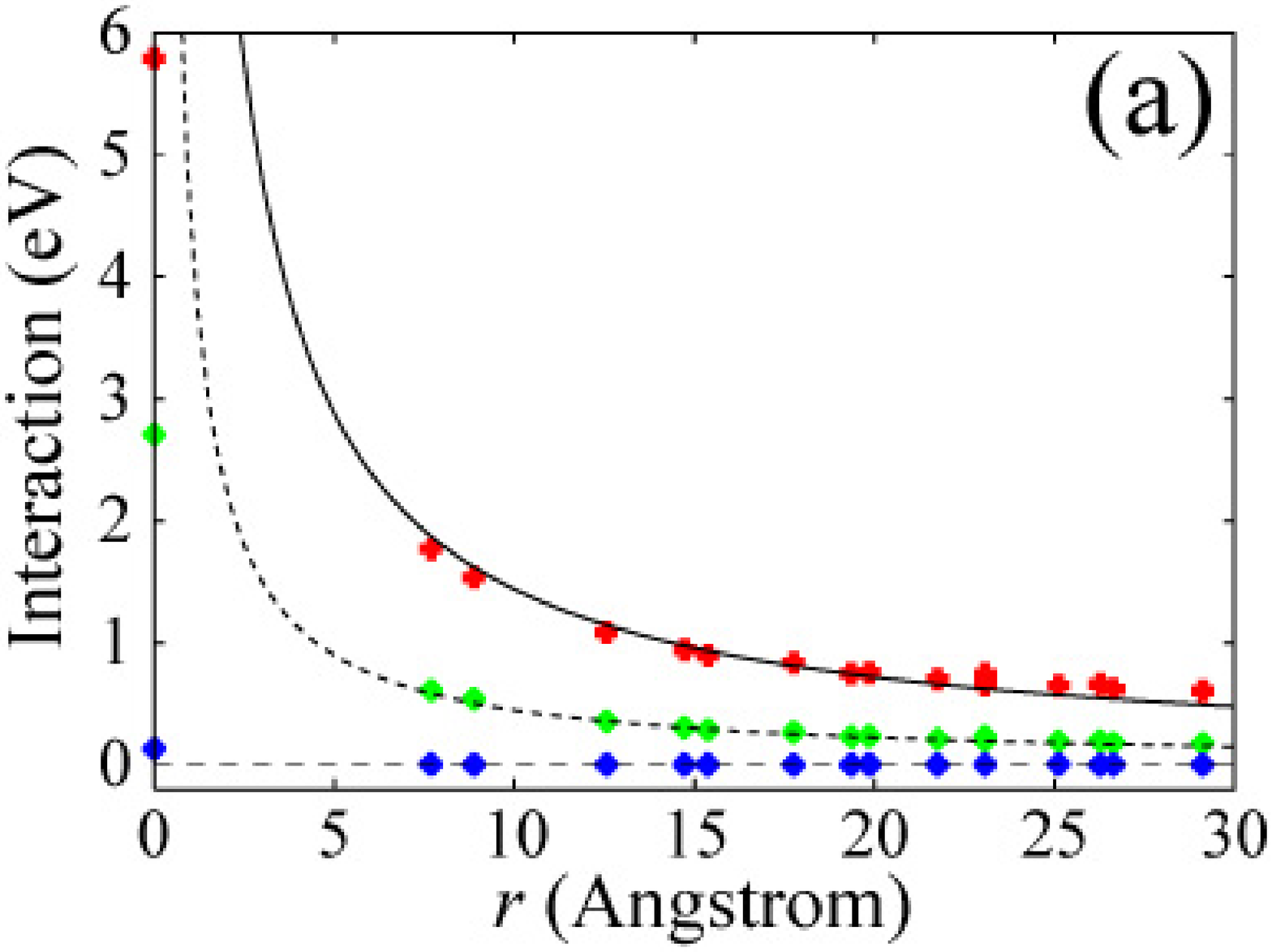} 
\includegraphics[width=0.35\textwidth]{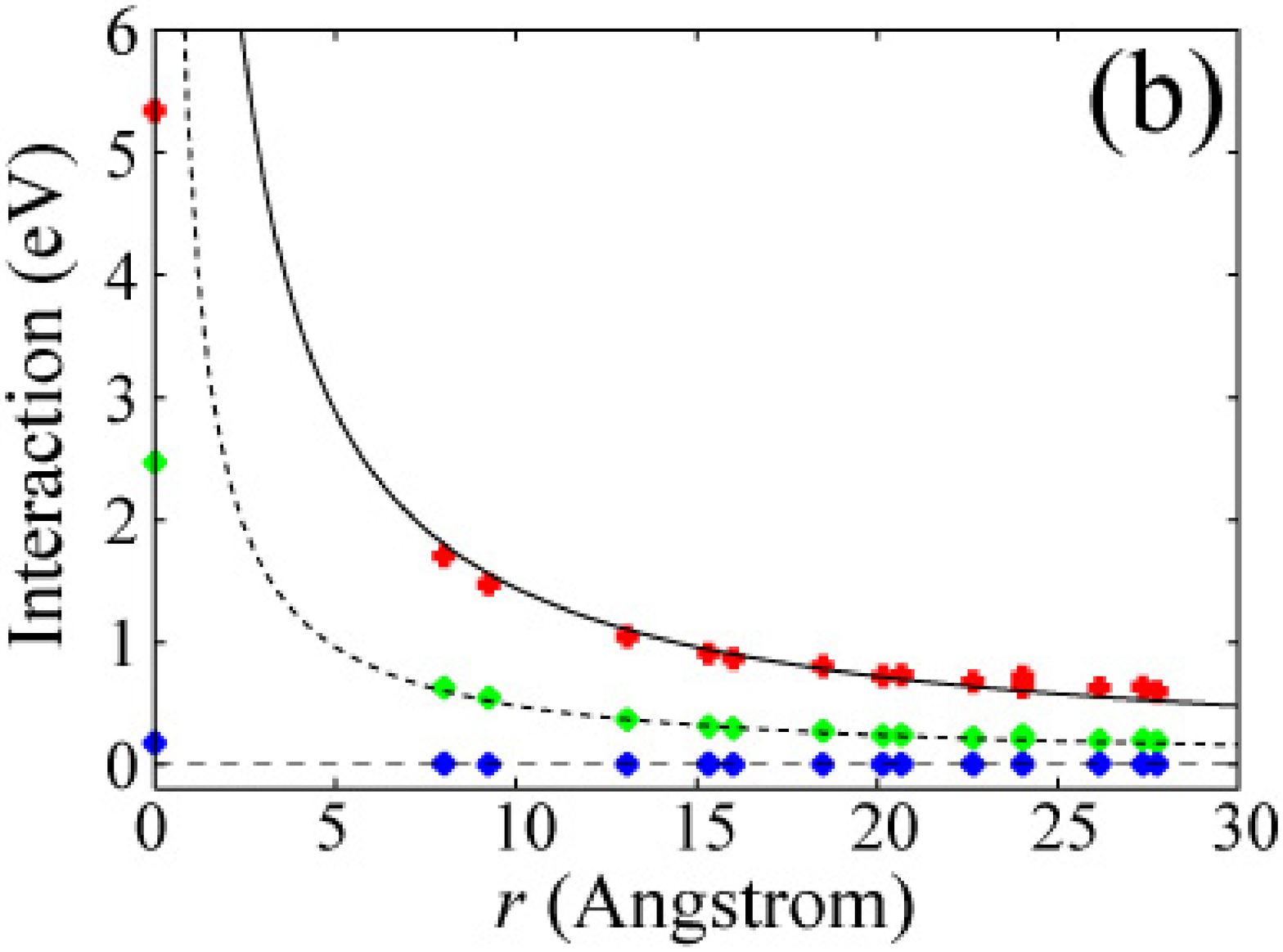} 
\caption{(Color online) Calculated screened Coulomb interactions of (a) sodium electrosodalite and (b) potassium electrosodalite as a function of the distance between the centers of maximally localized Wannier orbitals displayed in Fig.~\ref{fig:Wannier}. The red, green, and blue dots represent the result with the bare, constrained random phase approximation (cRPA), and full-RPA interactions, respectively. The solid and dotted curves denotes $1/r$ and $1/(\epsilon r)$, respectively, where $\epsilon$ is a macroscopic dielectric constant calculated with the cRPA.} 
\label{fig:Interaction}
\end{figure}

Figure \ref{fig:Interaction} plots the Wannier matrix elements of the screened Coulomb interaction $V_{ij}$ (green dots) calculated with constrained RPA, as a function of the distance between the centers of the MLWOs; $r=|\langle \phi_{i}| \br | \phi_{i} \rangle - \langle \phi_{j}| \br | \phi_{j} \rangle |$. 
The panels (a) and (b) show the results of the sodium electrosodalite and of the potassium electrosodalite, respectively. 
The $V_{ij}$ decays as an isotropic function of $1/(\epsilon r)$ (dotted line) where $\epsilon$ is a macroscopic dielectric constant calculated with cRPA.
 The value of $\epsilon$ is 3.2 for the sodium electrosodalite and 3.0 for the potassium electrosodalite. 
For comparison, we also plot bare Coulomb interactions (red dots), which should decay as $1/r$ (solid line) beyond the nearest-neighbor distance ($\ge$7 \AA). 
We see that the bare Coulomb interaction is reduced in less than half by considering the screening effect with cRPA. 
On the top of this, the full RPA screened Coulomb interactions are shown as blue dots, which are nearly zero, except for the onsite value at $r$ = 0. 
The exchange interactions of $J_{ij}$ were found to decay very quickly; the magnitude is nearly zero, except for the nearest and next-nearest values.
This quick decay was the same for the three cases of the bare, cRPA, and full RPA. 

We summarize in TABLE \ref{PARAM} the principal parameters in the resulting Hubbard model of Eqs.~(\ref{H_Hub})-(\ref{H_J}); the interactions up to the next nearest neighbors.  
The table compares $U$, $V_1$, $V_2$, $J_1$, and $J_2$ calculated with the three-type interactions of the bare, cRPA, and full RPA. 
We see that the calculated values become small in order of increasing the screening (bare $\to$ cRPA $\to$ RPA). 
From cRPA to full RPA, the diagonal parts of the Coulomb interaction, $U$, $V_{1}$, and $V_{2}$, are significantly reduced by more than one order of magnitude.
This is simply because a metallic screening process is switched on at the full RPA.
In contrast, the off-diagonal parts of the Coulomb interaction, $J_{1}$ and $J_{2}$, are not so screened.\cite{Miyake}
The kinetic-exchange values $K_1$ and $K_2$ obtained via Eq.~(\ref{K_ij}), in the bottom two in the table, exhibit increasing tendency with bare $\to$ cRPA $\to$ RPA. 

There are discernible differences between the sodium electrosodalite and the potassium electrosodalite; for example, for cRPA, $U$ of the sodium electrosodalite (2.71 eV) is somewhat larger than that of the potassium electrosodalite (2.47 eV) and $J_1$ of the sodium case (27.0 K) is nearly half of the potassium case (44.5 K).
These results can be consistently understood in terms of the smaller spatial spread of maximally localized Wannier orbitals of the sodium case (2.66 \AA) than that of the potassium case (2.91 \AA).

The accuracy of the second-order perturbation in deriving the exchange parameters in the Heisenberg model is checked by an estimate of $(U-V_1)/t_1$. 
The values are large enough for the both sodalites (in cRPA, 36.7 for the sodium electrosodalite and 27.1 for the potassium electrosodalite), indicating that the system is close to the atomic limit enough and the perturbation treatment is reasonably justified.
All odd-order contributions with respect to transfer $\mathcal{H}_t$ of Eq.~(\ref{H_t}) to the kinetic exchanges vanish, independent of the lattice.\cite{Takahashi}  
The first correction to the second-order perturbation arises from the fourth order, which is negligibly small. 

\begin{table}[h] 
\caption{List of interaction parameters in the single-band extended Hubbard model in Eq.~(\ref{H_Hub}), together with kinetic exchanges in Eq.~(\ref{K_ij}). The parameters obtained with the bare, constrained random phase approximation (cRPA), and full RPA are compared. Suffices ``1" and ``2" attached to $V$, $J$, and $K$ specify the nearest neighbors and the next-nearest neighbors, respectively. Units are eV for $U$, $V_1$, and $V_2$ and K for $J_1$, $J_2$, $K_1$, and $K_2$.} 

\ 

\centering 
\begin{tabular}{lr@{\ \ \ \ \ \ }r@{\ \ \ \ \ \ }r@{\ \ \ \ \ \	\ \ \ \ \ }r@{\ \ \ \ \ \ }r@{\ \ \ \ \ \ }r} \hline \hline
 & \multicolumn{3}{l}{sodium electrosodalite} & \multicolumn{3}{c}{potassium electrosodalite} \\ \hline
 & bare                    & cRPA                    & RPA                       & bare                    & cRPA                    & RPA \\ \hline \hline
   $U  $   & 5.79   & 2.71 & 0.13  & 5.34 & 2.47  & 0.17  \\ 
   $V_1$   & 1.77   & 0.61 & 0.01  & 1.70 & 0.63  & 0.01  \\ 
   $V_2$   & 1.54   & 0.54 & 0.00  & 1.47 & 0.54  & 0.00  \\ 
   $J_1$   & 56.9   & 27.0 & 22.0  & 97.2 & 44.5  & 39.0  \\ 
   $J_2$   & 22.4   & 10.6 & 8.6   & 20.9 & 9.9   & 8.1   \\ 
   $K_1$   & 18.9   & 36.3 & 596.5 & 29.4 & 58.2  & 631.8 \\ 
   $K_2$   &  5.6   & 11.0 & 182.5 &  5.8 & 11.7  & 129.2 \\ \hline \hline
\end{tabular} 
\label{PARAM} 
\end{table}

We show in TABLE \ref{PARAM_J} the theoretical Heisenberg exchanges ${\mathcal J}_1$ and ${\mathcal J}_2$ [Eq.~(\ref{J_eff})] obtained with the Hubbard-model parameters in TABLE \ref{PARAM} and compare those with the experimental values derived in Section~III. 
We see that the exchange couplings qualitatively change by considering the screening effect; the sign of the couplings changes from negative (ferromagnetic interaction) to positive (antiferromagnetic interaction) between the bare and cRPA. 
The values further enhance as proceeding from cRPA to full RPA but the latter gives a clear overestimate due to the large size of kinetic exchanges (see TABLE \ref{PARAM}). 
For the agreement between the theory and experiment, the cRPA is clearly the best among the three cases of the bare, cRPA, and full RPA. 
\begin{table}[h] 
\caption{List of parameters of the Heisenberg model in Eq.~(\ref{H_eff}), where $\mathcal{J}_1$ and $\mathcal{J}_2$ are the nearest and next-nearest exchange couplings. The theoretical values with the bare, constrained random phase approximation (cRPA), and full RPA are compared with the experimental results obtained in Sec.~III. The unit is K} 

\ 

\centering 
\begin{tabular}{lr@{\ \ \ }r@{\ \ \ }r@{\ \ \ }r@{\ \ \ }r@{\ \ \ }r@{\ \ \ }r@{\ \ \ }r} \hline \hline
 & \multicolumn{4}{c}{sodium electrosodalite} & \multicolumn{4}{c}{potassium electrosodalite} \\ \hline
 & bare & cRPA & RPA & Expt. & bare & cRPA & RPA & Expt. \\ \hline \hline
${\mathcal J}_1$ & $-$37.9 & 9.3 & 574.5 & 26 & $-$67.8 & 13.8 & 592.9 & 48 \\ 
${\mathcal J}_2$ & $-$16.8 & 0.4 & 173.9 &  8 & $-$15.1 &  1.8 & 121.1 & 20 \\ \hline \hline
\end{tabular} 
\label{PARAM_J} 
\end{table}

However, the calculated values of $\mathcal{J}$ with cRPA are still quantitatively underestimated from the experiment.  
This may be partially attributed to errors in the derived Hubbard-model parameters. 
A possible error is underestimation of the transfer parameters calculated as matrix elements of the Kohn-Sham Hamiltonian $\mathcal{H}_{{\rm KS}}$.
The $\mathcal{H}_{{\rm KS}}$ already includes the self-energy effect due to electron-electron interactions in the target bands of the Hubbard model as the exchange-correlation potential. 
As mentioned above, in the downfolding scheme,\cite{Aryasetiawan,Solovyev} this self energy must be excluded in the stage of the derivation of the Hubbard model.
If we use the exact $\mathcal{H}_{0}$ not including the target-band self energy, the magnitudes of the evaluated transfers will become larger quantitatively. 
We found that an artificial enhancement of the transfers by 20 \% leads to a satisfactory improvement to the underestimation of $\mathcal{J}$ observed above. 
 (For example, for the sodium electrosodalite, with this modification, the $\mathcal{J}_{1}$ and $\mathcal{J}_{2}$ values change from 9.3 K and 0.4 K to 25.3 K and 5.2 K, respectively.)

Another possibility of the error might arise in the interaction parameters evaluated by the constrained RPA.
The RPA leaves out the vertex correction in the polarization function. 
There are some studies in which the vertex correction is treated within local density approximation in density-functional framework.\cite{Louie,Northrup}
By considering this effect, the screening becomes larger. 
If we calculate the screened Coulomb interaction $W(\br,\br')$ with using this LDA dielectric function instead of the RPA one, we will obtain smaller values of the interaction parameters. 
We found that the use of an artificially smaller $U$ by 25 \% in a ${\mathcal J}$ estimation leads to an improvement; for sodium electrosodalite, we obtained $\mathcal{J}_{1}$ = 26.6 K and $\mathcal{J}_{2}$ = 5.4 K.
The quantitative discussions about the beyond RPA treatment are, however, not simple and need to be given more carefully in future studies.

Finally, we consider an effect of electron-lattice coupling on the results.
If an electron occupying a superatom localized $s$ orbital (see Fig.~\ref{fig:Wannier}) is transferred to the next site,  
one may then expect relaxation with an orbital expansion, leading to 
a reduction of the onsite Coulomb repulsion; with this expansion of the localized orbital, the excitation energy to the doubly-occupied state is reduced from $U - V_{ij}$ to $U - V_{ij} - \Delta S$, where $\Delta S$ is a stabilization energy due to the orbital expansion induced by a lattice deformation of tetrahedral cluster Na$_4$ confined in a $\beta$ cage.
Its energy scale can be the order $\sim$0.1 eV,\cite{Na3} and thus taking into consideration of this effect is expected to give a substantial improvement. 
The quantitative estimation of $\Delta S$ from first principles is, however, not so easy, which would require to solve technical issues including {\em ab initio} calculation for electron-lattice coupling.

\section{Conclusions}
To conclude, we have presented effective Heisenberg models describing the magnetic properties of alkali-cluster-loaded-sodalite systems. 
The derivation of the exchange couplings is based on the second-order perturbation to a single-band extended Hubbard model parameterized by {\em ab initio} density-functional and constrained RPA calculations. 
Main results in the present study are that (i) the direct-exchange couplings, dropped in the past studies, were estimated from first principles and were found to be the same energy scale as the kinetic exchanges and (ii) importance of considering the screening effect in the parameter derivation was found out; when the screening is properly considered, the net exchange couplings $\mathcal{J}_1$ and $\mathcal{J}_2$ become antiferromagnetic and the resulting exchange values are in a reasonable agreement with the experimental values on the order of ten K.

In this work, we have considered a single-band system; the low-energy electronic structures of the sodalite systems were captured in view of a superatom-$s$-electron picture.
It is interesting to apply the strategy presented here to other zeolites; for example, zeolites LTA and LSX described by multi-band systems. 
The cage size of these materials is bigger than that of the sodalite and many alkali atoms more than four can easily be doped. 
As a result, these materials will form partially-filled $p$-band structures of the superatoms. 
The minimal model of these systems are clearly the multi-band model, thus leading to a new intriguing magnetic property due to the Hund's rule coupling and/or its competition with the kinetic and direct exchanges. 
In fact, the temperature-dependence data of the magnetic susceptibility of the LTA zeolite strongly suggest the possibility of the highly nontrivial ferromagnetic ground state.\cite{Nakano} 
(The antiferromagnetic behavior suddenly changes to the ferromagnetic behavior at 50 K.)
There are active debates on this mechanism and {\em ab initio} calculations aiming at the construction of the effective Hamiltonians describing the low-energy physics of these systems will helpfully be contributed, which remains as future study.

\begin{acknowledgments}
We thank Professor Masatoshi Imada for suggesting detailed comparison between constrained RPA results and full RPA ones. 
We also acknowledge Professor Yasuo Nozue and Takehito Nakano for making their magnetic measurement data on potassium electrosodalite available to us and fruitful discussions. 
We thank Yoshihide Yoshimoto, Taichi Kosugi, Yoshiro Nohara, and Takashi Miyake for useful discussions. 
 This work was supported by a Grant-in-Aid for Scientific Research in Priority Areas, ``Development of New Quantum Simulators and Quantum Design'' (No. 17064004 and No. 19019012) of the Ministry of Education, Culture, Sports, Science and Technology (MEXT), Japan.
This research was partially supported by Scientific Research on Priority Areas of ``New Materials Science Using Regulated Nano Spaces" (No. 19051016) MEXT, Japan.
K.N. and R.A. acknowledge financial support from the Global COE Program ``the Physical Science Frontier," MEXT, Japan. 
All the computations have been performed on Hitachi SR11000 system at the Supercomputer Center, Institute for Solid State Physics, the University of Tokyo and on the same system of Supercomputing Division, Information Technology Center, the University of Tokyo.

\end{acknowledgments}



\begin{references}
\bibitem{Ref_PP3}
The sodium pseudopotential was generated in a slightly ionic (3$s$)$^{0.5}$(3$p$)$^{0.0}$(3$d$)$^{0.0}$ configuration by employing the cutoff radii $r_{3s}$ = $r_{3p}$ = $r_{3d}$ = 2.5 Bohr, while the potassium pseudopotential was constructed under the semicore configuration (3$s$)$^{2.0}$(3$p$)$^{6.0}$ for $r_{3s}$ = $r_{3p}$ = 2.6 Bohr.
The Al, Si, and O pseudopotentials were generated in the configurations of (3$s$)$^{2.0}$(3$p$)$^{0.5}$(3$d$)$^{0.0}$ for $r_{3s}$ = $r_{3p}$ = $r_{3d}$ = 2.0 Bohr, (3$s$)$^{2.0}$(3$p$)$^{1.5}$(3$d$)$^{0.0}$ for $r_{3s}$ = $r_{3p}$ = $r_{3d}$ = 1.8 Bohr, and (2$s$)$^{2.0}$(2$p$)$^{4.0}$ for $r_{2s}$ = $r_{2p}$ = 1.4 Bohr, respectively.
\end{references}
\end{document}